\documentclass{pasj00}
\SetRunningHead{M. Nobukawa et al.}
{Discovery of Neutral Lines from the Galactic Center Region}
\Received{}
\Accepted{}
\begin{document}
\title{Discovery of K-shell Emission Lines of Neutral Atoms in the Galactic Center Region}
\author{Masayoshi \textsc{Nobukawa}$^1$, Katsuji \textsc{Koyama}$^1$, 
Takeshi Go \textsc{Tsuru}$^1$, Syukyo G \textsc{Ryu}$^1$,
and Vincent \textsc{Tatischeff}$^2$}
\affil{$^1$Department of Physics, Graduate school of Science, Kyoto 
University, Sakyo-ku, Kyoto 606-8502}
\email{nobukawa@cr.scphys.kyoto-u.ac.jp}
\affil{$^2$Centre de Spectrom\'etrie Nucl\'eaire et de Spectrom\'etrie de
Masse, IN2P3-CNRS and Univ Paris-Sud, 91405, Orsay, France}

\maketitle

\begin{abstract}
The K-shell emission line of neutral irons from 
the Galactic center (GC) region is one of the key for the structure
and activity of the GC. The origin is still open question, but possibly
due either to X-ray radiation or to electron bombarding to neutral atoms.
To address this issue, we analyzed the Suzaku X-ray spectrum from the GC region 
of intense neutral iron line emission, and report on the discovery of
K$\alpha$ lines of neutral argon, calcium, chrome, and manganese atoms. 
The equivalent widths of these K$\alpha$ lines indicate
that the metal abundances in the GC region should be $\sim$1.6 and $\sim$4 of 
solar value, depending on the X-ray and the electron origins, respectively. 
On the other hand, the metal abundances in the hot
plasma in the GC region are found to be $\sim$1--2 solar.
These results favor that the origin of the neutral K$\alpha$ lines are due 
to X-ray irradiation.
\end{abstract}


\section{Introduction}
ASCA found clumpy structures of the 6.4~keV line
in the Sagittarius (Sgr) B2 and the Radio Arc regions (Koyama et al. 1996). 
Since the clumps correspond to giant molecular clouds, 
we refer to them as a "neutral clump" hereafter.

The X-ray spectrum of the Sgr B2 neutral clump has a prominent 6.4~keV 
line on the continuum emission with deep absorption of
$N_{\rm H}\sim10^{24}$~cm$^{-2}$ (Murakami et al. 2000). 
Recently, many new neutral clumps 
M\,0.74$-$0.09, M\,0.51$-$0.10 (Sgr~B1), G\,0.174$-$0.233,
M\,359.47$-$0.15, M\,359.43$-$0.07, M\,359.43$-$0.12, M\,359.38$-$0.00
have been discovered 
\citep{Ko07c, Yu07, No08, Fu09, Na09}.

The origin of a neutral clump, particularly the 6.4~keV line, 
is one of the open questions
among high energy phenomena in the Galactic center (GC) region. 
Koyama et al. (1996) and Murakami et al. (2000) suggested that 
the 6.4~keV emission from the neutral clumps is due to the K-shell 
ionization of iron atoms by external X-rays, possibly from 
the super-massive black hole, Sgr~A* (an X-ray reflection nebula; XRN). 
Although the present X-ray luminosity of Sgr~A* is 
$\sim10^{33-35}$~erg~s$^{-1}$~cm$^{-2}$
(Baganoff et al. 2001, 2003),
the luminosity required for the 6.4~keV emission is 
$10^{38-39}$~erg~s$^{-1}$~cm$^{-2}$. 
Therefore, Sgr~A* in a few hundred years ago 
would have been $10^{3-6}$ times brighter than now (e. g. Koyama et al. 1996).
On the other hand, Yusef-Zadeh et al. (2007) proposed that the origin
of the 6.4~keV emission would be low energy cosmic-ray electrons
($E_{\rm e}=$10--100~keV), because they found that the X-rays correlated with
non-thermal radio filaments.

Neutral atoms of lighter elements, such as Ar and Ca, should also exist in 
the neutral clumps, and hence would emit K-shell emission lines. 
These would provide new information to constrain the origin of the neutral clumps
in the GC region. 
However only the K-shell lines of neutral Fe and Ni atoms have been discovered so 
far (but see Fukuoka et al. 2009).

The Sgr~A region is the best place for the search of neutral 
K-shell lines of various elements because many bright neutral 
clumps have been 
found at ($l$, $b$)$\sim$(\timeform{0D.1}, \timeform{-0D.1}) 
\citep{Tsu99} and many observations have been performed with
Suzaku \citep{Ko07b, Hyo09}.

We re-analyzed the X-ray data obtained with the
X-ray Imaging Spectrometers (XIS; Koyama et al. 2007a) aboard 
Suzaku \citep{Mi07} and 
found emission lines from neutral Ar, Ca, Cr, and Mn atoms.
This paper reports on a detailed analysis, results, and discussion 
on the origin.


\section{Observations and Data Reduction}
\begin{table*}[t]
\caption{Observation Data List.}
\label{tab:obs_data}
\begin{center}
\begin{tabular}{ccccccc}
\hline
Observation ID & R.A. & Decl. & XIS & SCI & obs start & exposure time* \\ \hline
100027010 & 266.5146 & -28.9267 & 0\,1\,2\,3 & off & 2005-09-23 & 44.8~ks \\
100037040 & 266.5133 & -28.9266 & 0\,1\,2\,3 & off & 2005-09-30 & 43.0~ks \\
100048010 & 266.5135 & -28.9269 & 0\,1\,2\,3 & off & 2006-09-08 & 63.0~ks \\
102013010 & 266.5129 & -28.9278 & 0\,1\,3\ \ & on  & 2007-09-03 & 51.4~ks \\
\hline
  \multicolumn{2}{@{}l@{}}{\hbox to 0pt{\parbox{110mm}{\footnotesize
  \ 
    \par\noindent
	*After the data screening described in the text. 
    \par\noindent
  }}}
\end{tabular}
\end{center}
\end{table*}
The eastern vicinity of Sgr A* was observed 4 times in 2005, 2006, and 2007 
with Suzaku. Each pointing angle was almost 
the same. The observation information is listed in table~\ref{tab:obs_data}.

The XIS consist of three front-illuminated (FI; XIS\,0, 2, and 3) and 
one back-illuminated (BI; XIS\,1) CCD camera systems. All of the XIS are
placed on the focal planes of the X-Ray Telescopes (XRT; \cite{Se07}).
The size of the field of view is $\timeform{17'.8}\times\timeform{17'.8}$
and the half-power diameter is $\timeform{1'.9}$--$\timeform{2'.3}$.
XIS\,2 has been out of use since 2006 November, hence we 
did not use the XIS\,2 data on the 2007 September.

The XIS was working with the normal clocking and the 
full-window mode during all observations. The XIS pulse-height 
data for each X-ray event were converted to Pulse Invariant (PI) channels 
using the {\tt xispi} software version 2008-04-10, and the calibration database 
version 2008-08-25. We removed the data during the epoch of low-Earth 
elevation angles of less than 5$^{\circ}$ 
(ELV$<5^{\circ}$), day Earth elevation angles of less than 10$^{\circ}$
(DYE{\_}ELV$<10^{\circ}$), and the South Atlantic Anomaly.
The good exposure times are listed in table~\ref{tab:obs_data}. 

Since the XIS CCDs have been gradually degraded by on-orbit particle radiation,
the CCD performances have been restored by the Spaced-row charge injection (SCI)
technique since 2006 October \citep{Uchi08}. Thus, the data in 2007 September
were taken with the SCI technique.
The overall spectral resolutions (FI/BI) at 5.9 keV were 150/150, 175/185, and
140/175~eV (FWHM) for the 2005, 2006, and 2007 observations, respectively.

Since the non-X-ray background (NXB) depends on the geomagnetic cut-off 
rigidity (COR) \citep{Tawa2008NXB}, we sorted the NXB with the COR values,
using {\tt xisnxbgen}, from the night-Earth data released by the Suzaku XIS team.
The COR-sorted NXB was subtracted from the raw data with the same COR.

Since the relative gain of the FI sensors were well calibrated and 
the response functions were essentially the same, we co-added the FI data.
We also summed the spectra of the four observations to increase the statistics.

We analyzed the data using the software package HEASoft 6.6.1. 
For spectral fittings, we made XIS response files using {\tt xisrmfgen}, 
and auxiliary files using {\tt xissimarfgen}.


\section{Analysis and Results}

\subsection{6.4~keV line Image}
\begin{figure}[tp] 
\begin{center}
\FigureFile(80mm,50mm){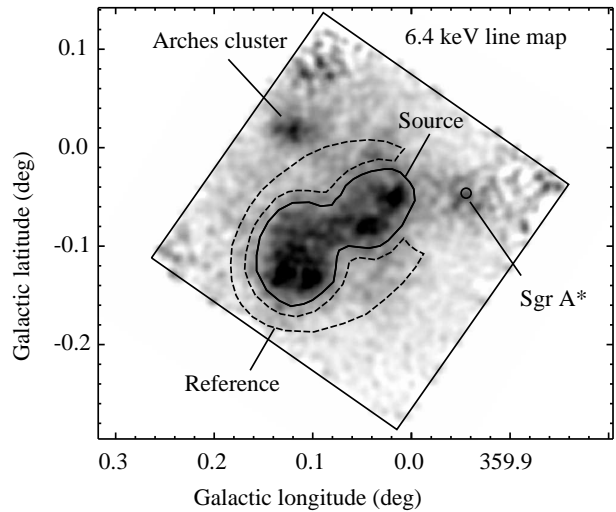}
\end{center}
\caption{X-ray image in the 6.4~keV neutral iron line (the 
underlying continuum fluxes are subtracted). The solid square 
is the field of view (FOV) of the XIS.
The small circle indicates the position of Sgr~A*.
We extracted the source spectrum from the solid region 
(Source). In order to examine the contribution
of the GC hot plasma, we also selected the surrounding region
(Reference) marked with the dashed line.
The areas of the Source and Reference regions are
43.3 and 40.5 arcmin$^2$, respectively.
The bright source in the northeast edge of the FOV
is the Arches cluster (e.g., Tsujimoto et al. 2007).}
\label{fig:image}
\end{figure}

In order to depict a neutral clump from the GC region, we made an X-ray image 
in the 6.4~keV line by the following procedures.
We first made X-ray images in the 5--6 keV and in the 6.3--6.5 keV bands, after 
subtracting the NXB and correcting the vignetting effect. 
We extracted the X-ray spectrum from a 8$'$ circular region near the field center,
and fitted the 5--6 keV band spectrum with a power-law model. 
The best-fit photon index in the 5.0--6.0~keV band is $1.25\pm0.07$.
The ratio of the photon flux in the 6.3--6.5~keV band 
to that in the 5.0--6.0~keV band was estimated to be 0.165, by extrapolation of the 
power-law index. We multiplied the 5.0--6.0~keV band image by 0.165, 
and subtracted from the 6.3--6.5~keV band image. The result is shown in figure~1.

We can see a bright region near the center of the 6.4~keV line image, as is shown with 
the solid line of a gourd-like shape (hereafter, the Source region). 
Around the Source region, we can see fainter emissions, which may be due
to unresolved miner neutral clumps.

The X-ray emission from the GC hot plasma contributes to the spectrum of 
the Source region.
In order to examine the contribution of the GC hot plasma accurately, 
we selected the dashed region, excluding the supernova
remnant, Sgr A East \citep{Ma02}, whose size and position are $\sim3'$
and ($l$, $b$)$\sim$(\timeform{-359D.5}, \timeform{-0D.5}).
Hereafter, we call this region as Reference. 


\subsection{Gain Tuning}
The nominal uncertainty of the absolute energy of XIS 
is $\sim10$~eV (the Suzaku XIS 
team\footnote{http://www.astro.isas.ac.jp/suzaku/doc/suzaku\_td}).
To be more accurate, we checked the energy of the K$\alpha$ lines of H-like 
S (S\emissiontype{XV}) of the spectra and the calibration 
source $^{55}$Fe (emits the Mn\emissiontype{I}~K$\alpha$ line at 
5895~eV) attached at the corner of the XIS. 
Compared with the theoretical value obtained from
Atomic \& Molecular Database in the Institute of Applied Physics 
and Computational Mathematics 
(CAMDB)\footnote{http:\/\/www.camdb.ac.cn\/e\/spectra\/spectra\_search.asp}, 
we found that the energy differences 
of S\emissiontype{XV} K$\alpha$ and Mn\emissiontype{I} K$\alpha$ were 
$5\pm2$ and $0\pm1$~eV for FI, while those for BI were consistent with
the theoretical energies.
We, therefore, fine-tuned the energy scale for FI as a liner function;
\begin{eqnarray}
\Delta E = -0.0015\times (E_0 - 5895)  {\rm\ (eV)},
\end{eqnarray}
where $E_0$ and $\Delta E$ are the original and energy shift, respectively.


\subsection{Model Construction}

\begin{figure}[t] 
\begin{center}
\FigureFile(80mm,50mm){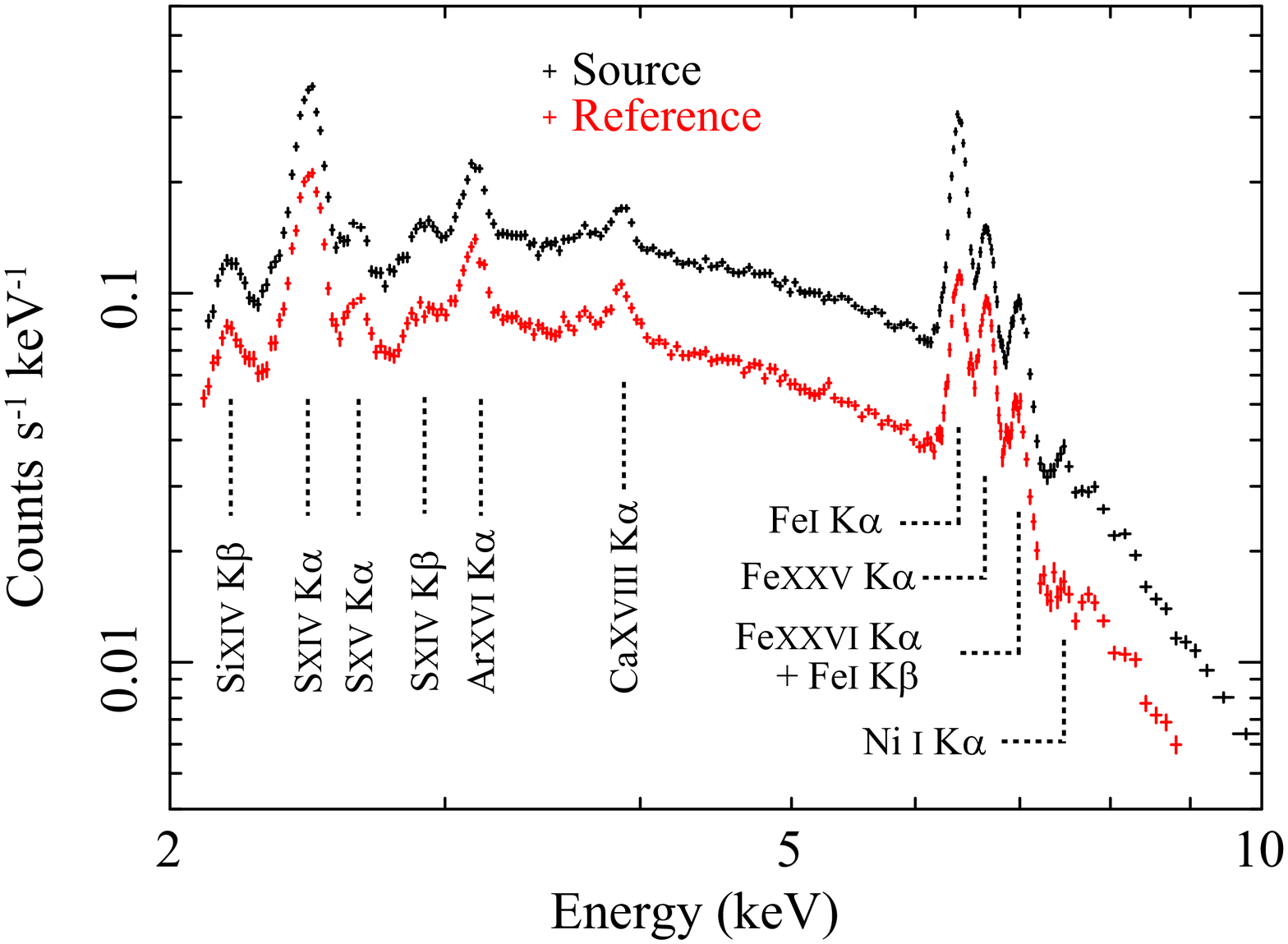}
\end{center}
\caption{X-ray Spectra (FI) in the Source 
(black) and the Reference (red) regions, where the NXB was already 
subtracted. Identified emission lines are indicated
by the dashed lines with their corresponding elements.
Errors of the data are estimated at the 1$\sigma$ confidence 
level.}
\label{fig:fitting}
\end{figure}

Figure~2 shows the X-ray spectra of the FI sensor in the Source and Reference regions. 
The spectra have several ionized emission lines of heavy elements, 
such as Si, S, Ar, Ca, and Fe. 
These lines would come from the GC hot plasma with the temperature 
of $kT\sim7$~keV \citep{Ko07b} and possibly $kT\sim$1~keV \citep{Ryu09}.

On the other hand, we can see prominent K$\alpha$ lines of neutral Fe 
and Ni at 6.4~keV and 7.5~keV in the spectra. 
The neutral lines indicate the existence
of a large amount of neutral Fe and Ni atoms in the regions. 
The neutral lines from the Reference region may be due
to many faint unresolved neutral clumps.

\begin{figure*}[t] 
\begin{center}
\FigureFile(150mm,100mm){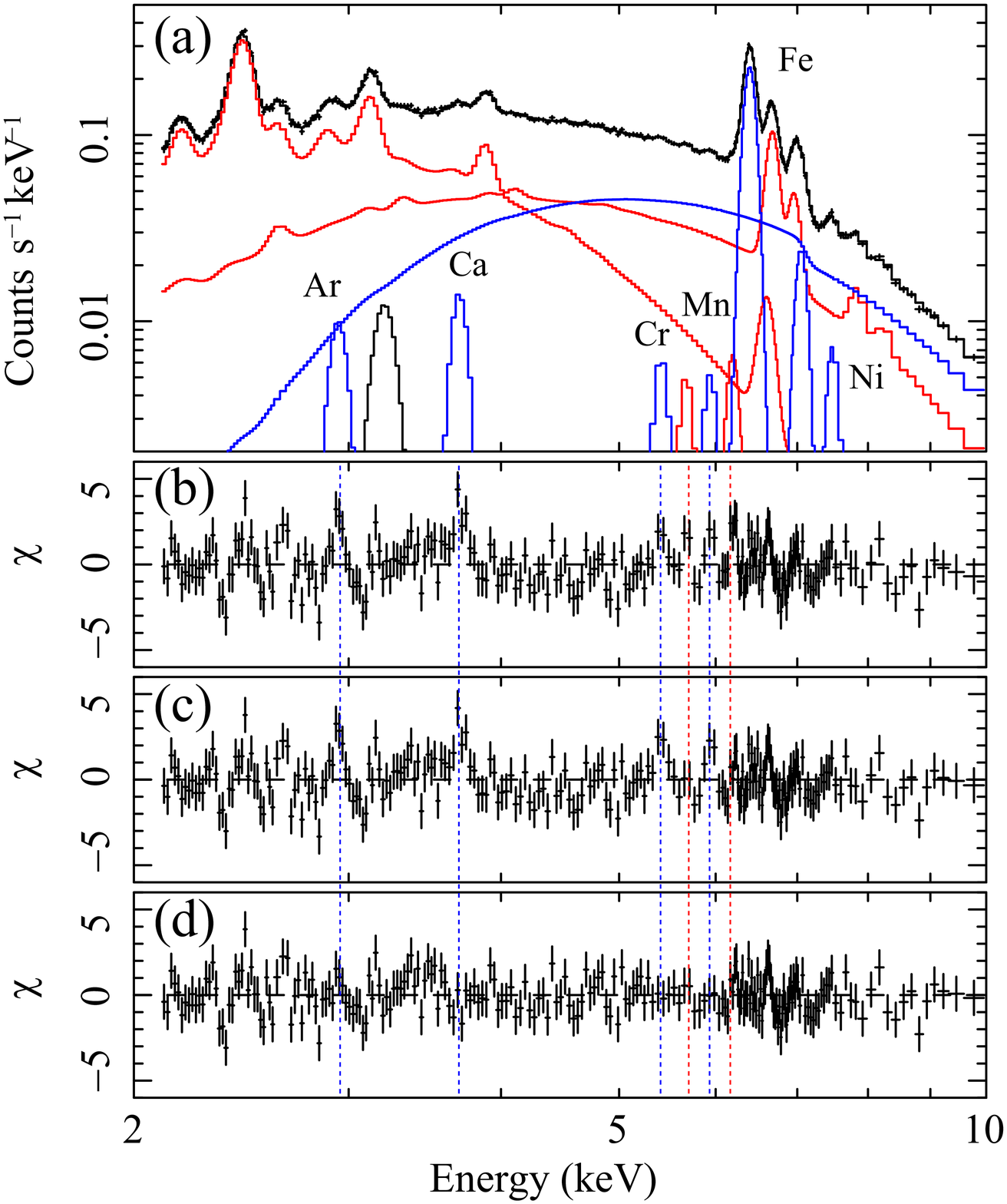}
\end{center}
\caption{(a): FI Spectrum of the Source region fitted with the 
final model (step 3) in subsection~3.4 (see text).
The red and blue histograms are the best-fit  $Plasma$ and $Neutral$ components,
respectively.
The Gaussian line at $3.2$~keV is added to compensate the uncertainty
of the instrumental response.
The blue and red dashed lines indicate the energies
of the neutral K$\alpha$ of Ar, Ca, Cr, Mn and the He-like
K$\alpha$ of Cr and Mn.
(b): Residual of the fitting with the models of 2-$kT$ $APEC$, 
a power-law, and neutral lines of Fe\emissiontype{I} 
K$\alpha$, K$\beta$, Ni\emissiontype{I} K$\alpha$ (step 1).
(c): Same as (b), but the ionized Cr and Mn lines were added (step 2). 
(d): Same as (c), but the neutral Ar, Ca, Cr, and Mn K$\alpha$
lines were added (step 3).
Errors of the data were estimated at the 1$\sigma$ confidence level.}
\label{fig:fitting}
\end{figure*}

The X-ray spectra of the two regions (Source and Reference)
would be a complex of the GC hot plasma and 
the neutral clumps \citep{Ko09, Ryu09}. 
We, therefore, made a fitting model composed of the GC hot plasma (here $Plasma$), 
the neutral clumps (hear $Neutral$) and
the Cosmic X-ray background (here $CXB$).
Thus, the model is given by;
\begin{eqnarray}
Model = Abs1 \times [Plasma + Neutral] + CXB, 
\end{eqnarray}
where $Abs1$ is absorption of the inter-stellar medium toward 
the GC region. 

\citet{Ryu09} found that the GC hot plasma ($Plasma$) has two-temperature components, 
and hence the $Plasma$ model can be described as
\begin{eqnarray}
Plasma = APEC1 + APEC2
\ ({\rm photons\ s}^{-1}\ {\rm cm}^{-2}),
\end{eqnarray}
where $APEC1$ and $APEC2$ are thin thermal plasma code in the
XSPEC package for the low ($kT_1$) and high ($kT_2$) temperature components, 
respectively.

$Neutral$ is composed of neutral lines and associated continuum emission, 
and hence can be given by
\begin{eqnarray}
Neutral = Abs2 \times [ A\times(E/{\rm keV})^{-\Gamma} 
+ Gaussians ]\nonumber \\
({\rm photons\ s}^{-1}\ {\rm cm}^{-2}),
\end{eqnarray}
where $Abs2$ is self-absorption in the neutral clumps.
Neutral lines are expressed by Gaussians, while the continuum is 
represented by a power-law model with the photon index $\Gamma$ and 
the normalization $A$. 

The cosmic X-ray background (CXB) flux is lower than the whole 
spectra shown in figure~2 by more than two orders of magnitude.
We nevertheless add the CXB model made by \citet{Ku02},
\begin{eqnarray}
CXB = Abs1\times Abs1 \times 7.4\times10^{-7} 
\times (E/{\rm keV})^{-1.41} \nonumber \\
({\rm photons\ keV}^{-1}\ {\rm s}^{-1}\ {\rm cm}^{-2}\ {\rm arcmin}^{-2}),
\end{eqnarray}
where absorption column density ($Abs1$) is applied twice
because the origin of the CXB is extragalactic.


\subsection{Spectral Fitting}

Since the surface brightness of $Plasma$
is not constant, the spectrum of $Neutral$ cannot be 
directly obtained by the subtracting the surrounding (Reference) region.
We therefore tried to simultaneous fit these two regions and 
to determine the spectra of the $Neutral$ and $Plasma$ models, separately.

The X-ray photons below $\sim$2 keV from the GC region are heavily 
absorbed by an inter-stellar medium of $N_{\rm H}\sim 6\times10^{22}$~cm$^{-2}$,
and the X-ray spectrum below $\sim$2 keV is dominated by the local
background \citep{Ryu09}. We therefore used only the 2--10~keV band for the 
spectral analysis.

In the fitting of $Plasma$, we assumed that the temperatures ($kT_1$ and $kT_2$)
and elemental abundances ($Z$) of Si, S, Fe, and Ni were common in the two regions 
(Source and Reference). Abundances of Ar and Ca were set to be the same as that of S.
Also, $Abs1$ was treated to be common in the two regions.
On the other hand, we set the normalizations of the hot plasma ($APEC1$, $APEC2$)
to be independent free parameters.

For $Neutral$, common free parameters among the two regions
are the power-law index ($\Gamma$), the equivalent 
widths of the neutral lines to the continuum, and line center energies.
The normalization of the power-law ($A$) is an independent free parameter.
With these constraints, we proceeded the model fitting with equation (2) 
along the following 3 steps: 

\ 

\subsubsection*{Step\,1: Gaussians for K-shell Lines from Fe and Ni}

For the $Neutral$ model of equation 4, we first tried a fitting with 
three Gaussians for Fe\emissiontype{I}~K$\alpha$ (6.40~keV), 
Fe\emissiontype{I}~K$\beta$ (7.06~keV), and Ni\emissiontype{I}~K$\alpha$ 
(7.49~keV) lines (e.g. \cite{Ko96, Ko07b}).

This fit left many residuals, including that at $\sim3.2$~keV. 
The energy is near the M-edge of Au (3.2--3.4~keV), and hence
the residuals are likely due to the calibration uncertainty of the 
response of the XRT. In fact, similar residuals were reported by 
\citet{Ku07}. We, thus, added a broad Gaussian line at $\sim 3.2$~keV 
according to \citet{Ku07}.
The best-fit center energy and the line width 
(1 sigma) of the Gaussian are $3.24\pm0.02$~keV and $0.07\pm0.02$~keV, 
respectively. 

The fitting residuals are displayed simply for the FI spectra of the 
Source region in figure~3b, although the fitting was made simultaneously 
for the both sensors (FI and BI) and both the two regions.
As shown in figure~3b, this model was not able to reproduce the 
observed spectra ($\chi ^2$/d.o.f is 1447/906).

\ 

\subsubsection*{Step\,2: K-shell Lines from Highly Ionized Cr and Mn}

One reason of the unacceptable fit is that the APEC model does not 
contain Cr and Mn K-lines at 5--6~keV.
We therefore added two Gaussian lines in the $Plasma$ model of equation (3) 
at $\sim5.7$ and 6.2~keV for the Cr\emissiontype{XXIII} and Mn\emissiontype{XXIV} K$\alpha$ 
lines, 
then, the center energies were obtained to be $5.68\pm0.02$ and $6.19\pm0.02$~keV, 
respectively. These are consistent with theoretical energies of $5.67$ and 
$6.17$~keV (CAMDB), the He-like Cr (Cr\emissiontype{XXIII}) and Mn 
(Mn\emissiontype{XXIV}).

\begin{figure}[t] 
\begin{center}
\FigureFile(80mm,50mm){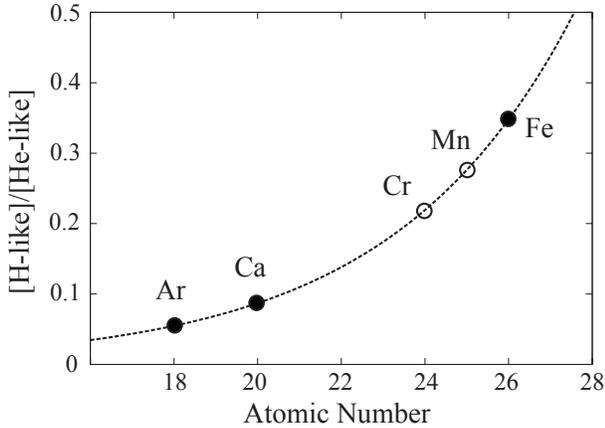}
\end{center}
\caption{Line-intensity ratios of He- and H-like ions.
The filled circles are calculated values using the APEC model.
The values of Ar, Ca, and Fe are 0.057, 0.085, and 0.35,
respectively. By interpolating these data, the line 
intensity ratios of Cr and Mn are estimated to be 0.22 
and 0.28 (open circles).}
\label{fig:lineratio}
\end{figure}

The hot plasma can also emit K$\alpha$ lines of
H-like Cr (Cr\emissiontype{XXIV}) and Mn (Mn\emissiontype{XXV})
at 5.94~keV and 6.44~keV, respectively.
Using the two-APEC model with the temperatures of 1 keV and 7keV
and the flux ratios given in table~\ref{fit_result}, 
we plotted the intensity ratio of H-like to He-like K$\alpha$ lines 
as a function of atomic number (Ar, Ca, Fe) in figure~\ref{fig:lineratio}.
By interpolation, we  estimated the line intensity ratios of Cr and Mn to be 0.22
and 0.28. We added these four lines in the $Plasma$ model:
\begin{eqnarray}
Plasma = APEC1 + APEC2 + 4\,Gaussians({\rm Cr,\ Mn}) \nonumber \\
\ \ ({\rm photons\ s}^{-1}\ {\rm cm}^{-2}), \ \ \ \ \ \ 
\end{eqnarray}
where the intensity ratios of H-like to He-like Cr and Mn were
fixed to be 0.22 and 0.28.
We also set the equivalent widths of Cr\emissiontype{XXIII}~K$\alpha$ 
and Mn\emissiontype{XXIV}~K$\alpha$ to the continuum of the $Plasma$ model
to be the same in the two regions. 

This fitting gave $\chi ^2$/d.o.f. of 1367/902. 
The residuals from this model in the Source spectrum are shown 
in figure~3c. In the next step (step 3), we used the $Plasma$ model of equation (6).


\ 

\subsubsection*{Step\,3: K-shell Lines from Neutral Ar, Ca, Cr, and Mn}

Since we still found line-like residuals at the energies 
of $\sim$3.0, 3.7, 5.4, and 5.9~keV, 
we added four Gaussian lines to the $Neutral$ model of equation (4) at these energies
(in figure~3c, also see dashed lines in figure~3). 
Then the line energies were found to be $2.94\pm0.02$, 
$3.69\pm{0.02}$, $5.41\pm0.04$, 
and $5.94\pm{0.03}$~keV, respectively.  
These values correspond to theoretical energies of 2.96, 3.69, 
5.41, and 5.90~keV for the K$\alpha$ lines of Ar\emissiontype{I}, 
Ca\emissiontype{I}, Cr\emissiontype{I}, and Mn\emissiontype{I}. 
The fitting residuals in the spectrum of the Source region 
are shown in figure~3d.

As demonstrated in figure~3d, the $\chi ^2$/d.o.f. is largely 
improved from 1367/902 (step 2)
to 1203/894 (step 3), but is not acceptable even at the 99\% confidence level.
Although the energy resolution in the hard X-ray band is well 
calibrated with the calibration source $^{55}$Fe (emits the 
Mn\emissiontype{I}~K$\alpha$ line), that in the soft band, especially
about 2~keV, may have some uncertainty due to the absence of available
calibration sources. 
Thus the deviation at $\sim$2.5~keV would come from the incomplete
response function in the soft energy band. 

Residuals are also found at $\sim$6.5--7.0~keV, with a slight
mismatch of the center energy of Fe\emissiontype{XXV}~K$\alpha$. 
This suggests that the GC hot plasma has more than two temperature 
components. We therefore tried a fitting with a 
three-$kT$ plasma model. Although the fit was improved from 
$\chi ^2$/d.o.f.$=$1203/894 to 1123/891, the best-fit 
parameters for the $Neutral$ and $Plasma$ models do not change from 
those in the two-$kT$ model within the statistical errors.
We, thus, regard that the two-$kT$ model after step 3 is a good 
approximation to derive the physical parameters.

For the discussion on the neutral clump in the Source region,
we list the best-fit parameters for the $Plasma$ model 
in table~\ref{fit_result}, while those for the $Neutral$ model 
are listed in table~\ref{result_neutral}. 
For comparison, the best-fit parameters of the Reference region are:
normalizations of $APEC1$ and $APEC2$ are $6.8\pm0.5$ and 
$0.90^{+0.06}_{-0.04}$ in the same unit shown in table~2, respectively.
$Abs2$ and normalization of the power-law component are 
$12.7^{+1.2}_{-1.8}\times10^{22}$ H~cm$^{-2}$ and 
$3.0^{+0.5}_{-0.4}\times10^{-3}$ photons keV$^{-1}$ s$^{-1}$ cm$^{-2}$, 
respectively.  
The other parameters are the same as the Source region.

\begin{table}[htbp]
\caption{Fitting results of $Plasma$.$^{*}$}
\label{fit_result}
\begin{center}
\begin{tabular}{lccc}
\hline
Component &Parameter  & Unit & Value \\ \hline
$Abs1$	& $N_{\rm H}$ & $10^{22}$~cm$^{-2}$ & $6.75\pm0.13$ \\
$APEC1$ 	& $kT_1$ & keV & $1.01^{+0.01}_{-0.02}$ \\
  	& Norm  & $\dagger$ & $12\pm1$ \\
$APEC2$ 	& $kT_2$ & keV & $7.0\pm0.1$ \\
  	& Norm  &  $\dagger$ & $1.3\pm0.1$ \\
Abundances &Si & solar & $2.52^{+0.09}_{-0.17}$ \\
	&S, Ar, Ca & solar & $1.87\pm0.07$ \\
	&Fe & solar & $1.16^{+0.07}_{-0.04}$ \\
	&Ni & solar & $1.64\pm0.37$ \\
\hline
\multicolumn{4}{c}{Gaussian Lines}\\
\hline
Cr\,{\small XXIII} K$\alpha$ 	& energy & (keV) & $5.68\pm0.02$  \\
				& EW & (eV) & $22\pm6$  \\
Mn\,{\small XXIV} K$\alpha$  	& energy &(keV) & $6.19\pm0.02$  \\
				& EW &(eV) & $39\pm6$  \\
\hline
  \multicolumn{2}{@{}l@{}}{\hbox to 0pt{\parbox{85mm}{\footnotesize
  \ 
    \par\noindent
    \footnotemark[*]The uncertainties indicate the 90\% confidence levels. 
    \par\noindent
    \footnotemark[$\dagger$] Emission measure 
    $10^{-12}/(4\pi D^2)\int n_{\rm e} n_{\rm H} dV$, 
    where $D$ is the distance to the source (cm), $n_{\rm e}$ and $n_{\rm H}$ 
    are the electron and hydrogen density (cm$^{-3}$), respectively.
    \par\noindent
  }}}
\end{tabular}
\end{center}
\end{table}

\begin{table*}[htbp]
\caption{Fitting Results of $Neutral$ and those of calculated values.$^{*}$}
\label{result_neutral}
\begin{center}
\begin{tabular}{cccccc}
\hline
\multicolumn{6}{c}{Neutral Lines} \\
Energy (keV) & Identification & Intensity$^{\dagger}$ & EW(eV)$^{\ddagger}$ & EW(XRN)$^\S$ & EW(LECRe)$^{\S}$ \\ 
\hline
$2.94\pm0.02$      & Ar\,{\small I} K$\alpha$ & $170^{+60}_{-40}$ & $140\pm40$ & 45 & 12 \\
$3.69\pm0.02$      & Ca\,{\small I} K$\alpha$ & $54^{+14}_{-9}$ & $83\pm13$ & 35 & 10 \\
$5.41\pm0.04$      & Cr\,{\small I} K$\alpha$ & $9.5\pm2.5$ & $24\pm7$ & 10 & 3.5 \\
$5.94\pm0.03$      & Mn\,{\small I} K$\alpha$ & $7.4\pm2.2$ & $22\pm7$ & 7.6 & 2.8 \\
$6.404\pm0.002$    & Fe\,{\small I} K$\alpha$ & $340\pm10$ & $1150\pm90$ & 730 & 270 \\
$7.06$ (fixed)     & Fe\,{\small I} K$\beta$  & $40\pm3$ & $160\pm20$ & 120 & 38 \\
$7.48\pm0.02$      & Ni\,{\small I} K$\alpha$ & $18\pm3$ & $83\pm13$ & 53 & 18 \\
\hline
\multicolumn{6}{c}{Continuum} \\
\hline
\multicolumn{2}{c}{Photon Index} & \multicolumn{2}{c}{$\Gamma$}  & \multicolumn{2}{c}{$1.87\pm0.04$}  \\ 
\multicolumn{2}{c}{Normalization} & \multicolumn{2}{c}{$\|$} & \multicolumn{2}{c}{$9.6^{+1.6}_{-1.3}$} \\
\multicolumn{2}{c}{$Abs2$($N_{\rm H}$)} & \multicolumn{2}{c}{$10^{22}$~cm$^{-2}$} & 
\multicolumn{2}{c}{$12.0\pm1.1$} \\ 
\hline 
\multicolumn{2}{@{}l@{}}{\hbox to 0pt{\parbox{130mm}{\footnotesize
  \ 
    \par\noindent
    \footnotemark[*]The uncertainties indicate the 90\% confidence levels. 
    \par\noindent
    \footnotemark[$\dagger$]
	Absorption-corrected line intensity in the unit of $10^{-6}$ photon s$^{-1}$ cm$^{-2}$.
    \par\noindent
    \footnotemark[$\ddagger$]
	Observed equivalent widths of the neutral lines to the power-law continuum in equation~4.
    \par\noindent
    \footnotemark[$\S$]
	Calculated equivalent widths in the XRN and LECEe scenarios.
    \par\noindent
    \footnotemark[$\|$]
	Normalization at 1~keV in the unit of $10^{-3}$ photons keV$^{-1}$ s$^{-1}$ cm$^{-2}$.
    \par\noindent
  }}}
\end{tabular}
\end{center}
\end{table*}


\section{Discussion}

\subsection{The GC hot plasma}
The absorption column density toward the GC region ($Abs1$) is
$\sim6.7\times10^{22}$ H~cm$^{-2}$, a typical value 
to the GC region (e.g., \cite{Mu04}). 
This is the first constrain that the low temperature 
$kT=1.01^{+0.01}_{-0.02}$~keV plasma is also in the GC region.

\citet{Ko07b}, using the same data set of this paper, reported 
the temperature of the GC hot plasma is $kT=6.5\pm0.1$~keV. 
The 1~keV plasma also emits Fe\emissiontype{XXV}~K$\alpha$,
but no significant Fe\emissiontype{XXVI}~K$\alpha$.
Since \citet{Ko07b} ignored the Fe lines from the 1.0~keV plasma,
they under-estimated the flux ratio of Fe\emissiontype{XXVI}~K$\alpha$ to 
Fe\emissiontype{XXV}~K$\alpha$, i.e., the plasma temperature.
Indeed, about 20\% of the Fe\emissiontype{XXV}~K$\alpha$ line may 
come from the $1.0$~keV plasma. The flux ratio of the Fe lines is estimated
to be $0.35$ and $0.42$ about the 6.5~keV and 7.0~keV plasmas by the APEC model,
respectively. Taking the contribution of the low temperature plasma into account,
the result in \citet{Ko07b} is consistent with our work.

The Fe and Ni abundances of 1.1--1.2 and 1.3--2.0 solar
are more precise than the previous work \citep{Ko07b}.
The abundances of lighter elements,  Si and S, (Ar, Ca), were respectively
determined to be 2.3--2.6 and 1.8--2.0 solar, for the first time.
The highly ionized Cr and Mn lines in the GC region were discovered 
for the first time, but details are beyond the scope of this paper.

\subsection{Origin of the Neutral Clump}

We discovered the K-shell lines of neutral Ar, Ca, Cr, 
and Mn from the bright neutral clump toward the Sgr~A region 
at the significance levels 
of 6.8, 9.6, 6.1, and 5.4~$\sigma$, respectively.
The absorption
column densities, $N_{\rm H}$($Abs2$) in the Source and 
Reference regions are $12.0(\pm1.1)\times10^{22}$ and 
$12.7^{+1.2}_{-1.8}\times10^{22}$ cm$^{-1}$, respectively.

The photon index, $\Gamma$ of the continuum (power-law)
is $1.87\pm0.04$, which is consistent with the result of
\citet{Ko09} in the same region.
The equivalent widths of the neutral K$\alpha$ lines to the power-law 
continuum are $\sim$140, 83, 24, 22, 1150, 83~eV in Ar, Ca, Cr, Mn, 
Fe, and Ni, respectively.

\begin{figure}[t] 
\begin{center}
\FigureFile(80mm,50mm){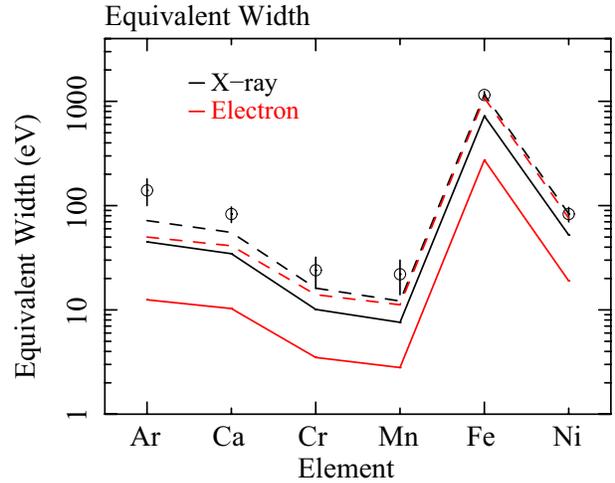}
\end{center}
\caption{
Equivalent widths of K$\alpha$ line of various neutral atoms.
Black and red lines are the calculated values for the X-ray (XRN) and 
electron (LECRe) scenarios, respectively. 
The data points marked with the open circles 
are the observed value in our work. Errors were estimated at the 90\% 
confidence level. The black and red dashed 
lines are to guide eyes, which are XRN and LECRe scenarios 
in 1.6 solar and 4.0 solar abundances, respectively.
}
\label{fig:EW}
\end{figure}

The major possibility for the origin of the neutral clumps
is the ionization of neutral atoms by either low energy
cosmic-ray electrons (LECRe: \cite{Yu07}) with an energy of 10--100~keV,
or X-rays of external sources (XRN: \cite{Ko96, Mu00}).
The ionization cross-sections for these processes are very different.
They also produce a continuum emission:
the bremsstrahlung in the LECRe scenario and Thomson scattering in the
XRN scenario. 
As a result, the two scenarios make different X-ray spectra.
In particular, sharp contrasts are the photon index of the continuum
and the equivalent widths of the neutral lines.

The continuum emission in the X-ray spectrum produced by the LECRe scenario
is an integration of the bremsstrahlung by electrons with various energies.
According to \citet{Ta03}, we calculated the X-ray spectrum with various 
indexes of their energy distributions into a molecular cloud. 
As a result, the photon index of $\Gamma=1.9$ 
corresponds to an index of the LECRe source
spectrum of $\alpha\sim3.0$. On the other hand, the Thomson 
scattering do not change
the photon index from the incident spectrum of the external source.

The equivalent width of the line is a good indicator for constraint of the origin.
According to the calculation in \citet{Ta03}, the equivalent widths produced
by the LECRe with the index $\alpha=3.0$ were estimated as shown with the red solid 
line in figure~5. Here, the elemental abundances in the neutral clump were 
assumed to be solar.
On the other hand, \citet{Mu00} estimated the XRN spectrum by a numerical simulation.
We improved the simulation \citep{Mu00} to the other relevant elements.
These are listed in table~3 and plotted in figure~5 together with the best-fit results.

From figure 5, the equivalent width of each element in the solar abundance
is larger than that expected in the both of the LCREe and XRN scenarios.
For the LCREe scenario, the abundances of the molecular cloud must
be $\sim$4-times larger than the solar value, while the XRN scenario 
requires $\sim$1.6 solar abundance. 

Since the molecular cloud may be formed by condensation of the ambient materials,
the abundances should be similar to those of 1--2 solar in the GC hot plasma. 
Accordingly, the neutral lines from the GC region likely come from the 
fluorescence by external X-rays. The photon index of 
$\Gamma\sim1.9$ is similar to the other neutral clumps in the Sgr~B and C
regions (Koyama et al.2007c; Nobukawa et al. 2008; Nakajima et al. 2009), 
which suggests that the irradiating source is a single object, possibly a 
super-massive black hole, Sgr~A*(e.g., \cite{Mu00}).

\section{Summary}
The summary of this work is as follows:
\begin{itemize}
	\item{We found that a $kT=1.0$~keV plasma exists in the GC region
	as well as the $kT=7.0$~keV plasma. The elemental abundances of Si,
	S (Ar, Ca), Fe, Ni were measured to be 2.3--2.6, 1.8--1.9, 1.1--1.2, 
	1.3--2.0 times larger than the solar value. We also discovered
	He-like Cr and Mn K$\alpha$ lines from the GC region.
	}
	\item{K-shell lines of neutral Ar, Ca, Cr, and Mn were firstly 
	discovered in addition to those of Fe and Ni. Equivalent widths 
	of these atoms are $\sim$140, 83, 24, 22, 1150, and 83~eV, 
	respectively.
	}

	\item{The observed equivalent widths of Ar, Ca, Cr, Mn, Fe and Ni
       favor the X-ray radiation origin to the molecular cloud with 1--2 
	solar abundance.
	}

	\item{The power-law index of $\Gamma=1.9$ for the continuum emission is similar to 
	the other neutral clumps, suggesting a single source origin, 
	possibly the super-massive black hole Sgr A*.
	}

\end{itemize}

\bigskip

The authors thank H Matsumoto and H Uchiyama for their comments.
This work is supported by the Grant-in-Aid for the Global COE Program 
"The Next Generation of Physics, Spun from Universality and Emergence" 
from the Ministry of Education, Culture, Sports, Science and
Technology (MEXT) of Japan.
This work is also supported by Grant-in-Aids from the Ministry of Education, 
Culture, Sports, Science and Technology (MEXT) of Japan, 
Scientific Research A, No. 18204015 (KK), 
and Scientific Research B, No. 20340043 (TT).
MN is supported by JSPS Research Fellowship for Young Scientists.


\begin{thebibliography}{}
\bibitem[Baganoff et~al.(2001)]{Ba01}
  Baganoff, F. K., et al. 2001, \nat, 413, 45
\bibitem[Baganoff et~al.(2003)]{Ba03}
  Baganoff, F. K., et al. 2003, \apj, 591, 891
\bibitem[Fukuoka et~al.(2009)]{Fu09}
  Fukuoka, R., Koyama, K., Ryu, S. G., \& Tsuru, T. G. 2009, \pasj, 61, 593
\bibitem[Inui et~al.(2009)]{In09}
  Inui, T. et al. 2009, \pasj, 61, S241 
\bibitem[Ishisaki et al.(2007)]{Is07}
  Ishisaki, Y., et al. 2007, \pasj, 59, S113
\bibitem[Hyodo et al.(2009)]{Hyo09}
  Hyodo, Y., Ueda, Y., Yuasa, T., Maeda, Y., Makishima, K., \& Koyama, K.
  2009, \pasj, 61, S99
\bibitem[Koyama et~al.(1996)]{Ko96}
  Koyama, K., Maeda, Y., Sonobe, T., Takeshima, T., Tanaka, Y., \&
  Yamaichi, S. 1996, \pasj, 48, 249
\bibitem[Koyama et~al.(2007a)]{Ko07a}
  Koyama, K., et al. 2007a, \pasj, 59, S23  
\bibitem[Koyama et al.(2007b)]{Ko07b}
  Koyama, K., et al. 2007b, \pasj, 59, S245 
\bibitem[Koyama et~al.(2007c)]{Ko07c}
  Koyama, K., et al. 2007c, \pasj, 59, S221 
\bibitem[Koyama et~al.(2009)]{Ko09}
  Koyama, K., Takikawa, Y., Hyodo, Y., Inui, T., Nobukawa, M.,
  Matsumoto, H., \& Tsuru, G. T. 2009, \pasj, 61, S255 
\bibitem[Kubota et~al.(2007)]{Ku07}
  Kubota, A. et al. 2007, \pasj, 59, S185 
\bibitem[Kushino et~al.(2002)]{Ku02}
  Kushino, A., Ishisaki, Y., Morita, U., Yamasaki, N. Y., Ishida, M.,
  Ohashi, T., \& Ueda, Y., 2002, \pasj, 54, 327 
\bibitem[Maeda et al.(2002)]{Ma02} 
  Maeda, Y., et al, \apj, 570, 671
\bibitem[Mitsuda et~al.(2007)]{Mi07}
  Mitsuda, K., et al. 2007, \pasj, 59, S1
\bibitem[Muno et~al.(2004)]{Mu04}
  Muno, M.~P., et al.\ 2004, \apj, 613, 326 
\bibitem[Murakami et~al.(2000)]{Mu00}
  Murakami, H., Koyama, K., Sakano, M., Tsujimoto, M., \& Maeda, Y. 2000, \apj, 534, 283
\bibitem[Nakajima et~al.(2009)]{Na09}
  Nakajima, H. et al. 2009, \pasj, 61, S233 
\bibitem[Nobukawa et~al.(2008)]{No08}
  Nobukawa, M. et al. 2008, \pasj, 60, S191 
\bibitem[Ryu \etal (2009)]{Ryu09}
  Ryu, G. S., Koyama, K., Nobukawa, M., Fukuoka, R., \& Tsuru, G. T., 2009, \pasj, 61, 751
\bibitem[Serlemitsos et al.(2007)]{Se07}
  Serlemitsos, P., et al. 2007, \pasj, 59, S9
\bibitem[Tatischeff(2003)]{Ta03}
  Tatischeff, V. 2003, in Final Stage of Stellar Evolution, ed. C. Motch 
  \& Hameury (EAS publication Series vol.7), 79 (astro-ph/0208397v1)
\bibitem[Tawa et al.(2008)]{Tawa2008NXB}
  Tawa, N., et al. 2008, \pasj, 60, S11
\bibitem[Tsuboi et al.(1999)]{Tsu99}
  Tsuboi, M., Handa, T., \& Ukita, N. 1999, \apjs, 120, 1
\bibitem[Tsujimoto et~al.(2007)]{Tsu07}
  Tsujimoto, M., Hyodo, Y., \& Koyama, K. 2007, \pasj, 59, S229 
\bibitem[Uchiyama et al.(2009)]{Uchi08}
  Uchiyama, H., et al. 2009, \pasj, 61, S9
\bibitem[Yusef-Zadeh et al.(2007)]{Yu07}
  Yusef-Zadeh, F., Muno, M., Wardle, M., \& Lis, D. C. 2007, \apj, 656, 847
\end{thebibliography}
\end{document}